\newcommand{\cL}{{\cal L}}            %
\newcommand{\cN}{{\cal N}}        %
\newcommand{\cV}{{\cal V}}        %

\newcounter{sentence}
\newcounter{scount}

\newcommand{\join}{\mbox{$\cup$}}        
\newcommand{\lmean}{\mbox{${\bf [}\! [$}}  
\newcommand{\rmean}{\mbox{${\bf ] \! ]}$}}

\newtheorem{myexp}{Example}

\documentclass[12pt]{fundam}
\usepackage{proof}

\begin{document}
\title{A generalization of falsity in finitely-many valued logics}
\address{Nissim Francez,
Department of Computer Science,
Technion -- Israel Institute of Technology, 
Haifa 32000, Israel.
Email: \texttt{francez{@}cs.technion.ac.il}.
}

\author{Nissim Francez}
\maketitle
\runninghead{Nissim Francez}{A generalization of falsity in finitely-many valued logics}
\section{Introduction}
In propositional classical logic, if a formula $\varphi$, under some valuation $\bf v$,
is not true, then $\varphi$  is false, and if it is not false, it is true.
 This toggling between truth and falsehood is captured in
 propositional classical logic by means of {\em negation} `$\neg$':, with its  truth-table':
\[
\begin{array}{c|c}
\varphi & \neg \varphi\\
-- & --\\
t &f\\
-- & --\\
f & t
\end{array}
\]
Consider now  some multi-valued\footnote{I consider here only finitely many
 truth-values.} logic $\cL$ with a set of truth-values\footnote{I ignore here the issue, orthogonal to
 our interest, whether $\cV$ is a set of truth-values or
 merely some {\em semantic values}.} 
 \[
 \cV=\{v_{1},\cdots,v_{n}\},\ n \ge 2
 \]
 {\em {\bf Q}: What does it mean that under some valuation ${\bf v}$, some $\varphi$
{\em  does not} have the truth-value $v_{i}$ for some $v_{i} \in \cV$}?\newline
{\em And, in particular, can this meaning be captured by means of a suitable negation in $\cL$}?\newline
Suppose we have already identified $v_{1}$ with $t$ and $v_{n}$ with $f$
(see \cite{FK18a} for one such identification; see Section {\ref{sec:unique} for
the  definition used where). Then, there is interest
in the special instances of the question {\bf Q}:\newline
What does it mean that $\varphi$ is not true 
under some valuation {\bf v}, or not false.

Traditionally in multi-valued logics,
negation is viewed {\em (truth) functionally}, $\neg \varphi$ mapping the truth-value $v_{i}$ of $\varphi$
to some other, {\em specific}, truth-value $v_{j}$, where $i=j$ is not excluded\footnote{See $K_{3}$ below for a case of $i=j$.}. 
This mapping is again depicted as a multi-valued truth-table.
Some well-know examples are listed below, without considerations of interpretation
of those truth-values.
\begin{description}
\item[Kleene's  $K_{3}$ \cite{Kl52}:]
Here $\cV=\{t, n, f \}$, and the truth-table for negation is
\[
\begin{array}{c|c}
\varphi & \neg \varphi\\
-- & --\\
t &f\\
-- & --\\
n &n\\
-- & --\\
f & t
\end{array}
\]
\item[ Belnap-Dunn  {\em first-degree entailment (FDE)}
 \cite{Be76,Be77,Du76}:]
 Here $\cV=\{t, b, n, f \}$, and the truth-table for negation is
\[
\begin{array}{c|c}
\varphi & \neg \varphi\\
-- & --\\
t &f\\
-- & --\\
n &n\\
-- & --\\
b &b\\
-- & --\\
f & t
\end{array}
\]
\item[Post cyclic negation \cite{Pos21}:]
$\cV=\{v_{0},\cdots,v_{n-1}\}$ and negation is {\em cyclic}.
\[
\begin{array}{c|c}
\varphi & \neg \varphi \\
--- &  ---\\
v_{0}   & v_{1}\\
--- &  ---\\
v_{1}  & v_{2}\\
--- &  ---\\
\vdots  & \vdots\\
--- &  ---\\
v_{n-1} & v_{0}
\end{array}
\]
\end{description}
Consequently, the question {\bf Q} is traditionally answered as follows: for any $v_{i} \in \cV$, if,
under a valuation {\bf v}, $\varphi$ does not have 
truth-value $v_{i}$, then $\varphi$ has under {\bf v} some
{\em specific}  truth-value $v_{j}$,  where $j=i$ is not excluded.\newline

In this paper, I aim at another way to answer the question {\bf Q}:
{\em If, under some valuation {\bf v}, $\varphi$ {\em does not} have truth-value $v_{i}$,
this is understood as $\varphi$ having,  under ${\bf v}$, {\em non-deterministically},
{\em any other} (not functionally determined) truth-value in $\cV$}.\newline
There is no ``privileged' $v_{j}$  materializing not having 
the value $v_{i}$!

This means that traditional multi-valued negations, as exemplified above, cannot be used
to express this interpretation of not having $v_{i}$.
Instead, I introduce another operator, that generalizes negation in multi-valued logics
as a non-deterministic operator. 
To distinguish our approach, I use a unary operator `$N$' instead of `$\neg$'.

I consider $n$ operators, $N_{i}$, for $1 \le i \le n$.
The intended meaning of $N_{i} \varphi$, when true under some valuation {\bf v}, 
is that $\varphi$  {\em does not have the truth-value} $v_{i}$ under {\bf v}. 
This, however, is not taken to mean as having some
{\em specific} truth-value $v_{j}$; rather, it is taken to mean that
$\varphi$ has, {\em non-deterministically}, {\em any} value different from $v_{i}$.

 Thus, $N_{i} \varphi$ can never (i.e., for no valuation ${\bf v}$)
 share the same truth-value with $\varphi$. It reflects the
 meta-linguistic negation of `$\varphi$ has truth-value $v_{i}$'. In this, $N_{i}$ differ  from
 $\neg \varphi$ in traditional multi-valued logics, where 
 ${\bf v} \lmean \varphi \rmean = {\bf v} \lmean \neg \varphi \rmean$ is certainly possible, e.g., for the truth-value $n$ in $K_{3}$ as shown
  above.

As for the intended meaning of $N_{i} \varphi$  when having a truth-value $v_{j} \ne v_{1}= t$,
this will be specified below once the theory is set up.
 
 As the framework for our study, we chose {\em located sequents}, introduced and studied in general
 in \cite{KF17b}, and used for a related issue in \cite{FK17a}.
 The formalism is delineated in Section  \ref{sec:located}.
 \section{Preliminaries: located formulas and sequents}
 \label{sec:located}
For $n\ge 2$, let $\cV=\{v_{1},\cdots,v_{n}\}$  be a collection of
 truth-values underlying a
multi-valued  logic $\cL^{n}$ with a propositional  object-language $L_{n}$  with, possibly,  some additional unspecified connectives defined by truth-tables over $\cV$. 
Let $\hat{n}=\{1,\cdots,n\}$.
\begin{definition}[located formulas\footnote{This is special case
of what is known in the literature as a signed formula.}]
 A {\em located formula}
($l$-formula)  is a pair $(\varphi, k)$, where $\varphi$ is an
object-language  formula and $k \in \hat{n}$. We say
that $(\varphi, k)$ {\em locates} $\varphi$ at $v_{k}$.
\end{definition} 
The intended interpretation
of $(\varphi, k)$ is that $\varphi$ is associated with 
  the truth-value  $v_{k} \in \cV$.
\begin{definition}[located sequents]
A {\em located sequent} ($l$-sequent) $\Pi$ has the form 
$\Gamma : \Delta$,
where $\Gamma,\ \Delta$ are (possibly empty) finite collections\footnote{The exact nature
of a collection, e.g., a set or a multi-set, depends on the specific logic being defined.}
of $l$-formulas. 
\end{definition}
I use ${\bf \Pi}$ for sets of $l$-sequents.
Let $\sigma$ range over valuations, mapping formulas to truth-values
in $\cV$;
for atomic sentences the mapping is arbitrary, and it is extended to compound formulas so as to respect
the truth-tables of the operators. Below, I define the central semantic notions as applicable
to $l$-sequents.
\begin{definition}[satisfaction, consequence]
\begin{description}
\item
\item[{\bf satisfaction}:]
$\sigma \models \Pi ( =\Gamma : \Delta)$ iff:
\begin{equation}
 {\rm if}\ \sigma \lmean \varphi \rmean=v_{k}\ {\rm for\ all}\ (\varphi, k) \in \Gamma,\ {\rm then}\ \sigma \lmean \psi \rmean=v_{j}\ {\rm for\ some}\ (\psi, j) \in \Delta
\end{equation}
\item[{\bf consequence}:]
\begin{equation}
{\bf \Pi} \models \Pi\ {\rm iff\ for\ every}\ \sigma :\ \sigma \models \Pi^{\prime}\ {\rm for\ all}\ \Pi^{\prime} \in {\bf \Pi}\
{\rm implies}\ \sigma \models \Pi
\end{equation}
\item[{\bf validity}:] $\Pi$ is {\em valid} iff $\sigma \models \Pi$ for every $\sigma$.
\end{description}
\end{definition}
In \cite{FK17a,KF17b}, 
 various  proof-systems over $l$-sequents are presented
 (in a different notation)
sound and (strongly) complete for 
the above consequence relation, constructed from the
truth-tables in a uniform way. The multi-valued ND-systems
$\cN^{n}$ (over $l$-sequents) with their structural and logical rules for an arbitrary 
$n$-ary connective are presented in an appendix.
\section{Transparent falsity and binary poly-sequents}
\label{sec:bivalent}
\subsection{Transparent falsity and disquotation}
As a preliminary step, I consider the case where\footnote{For
better readability, I use $\{t,f\}$ instead of $\{v_{1},v_{2}\}$.} $n=2$, in which the non-determinism
involved  is only apparent, since `any truth-value other than $t$' is just $f$, 
and `any truth-value other than $f$' is just $t$. 
This section is an adaptation from \cite{FK18a}.

Suppose we want to add to classical logic a {\em transparent falsity-predicate} $F(x)$.
What would be the way to express falsity? 
Fortunately, because of the properties of
classical negation, where the truth of $\neg \varphi$ expresses the falsity of $\varphi$,
we can use it for creating such an analog to the disquotation property of the well-known
truth predicate:
\begin{equation}
\label{eq:falsity}
(DF)\ \ 
F(\hat{\varphi}) \leftrightarrow \neg \varphi
\end{equation}
where $\hat{\varphi}$ is a name for $\varphi$ (e.g., the G$\ddot{o}$del number).
The transparency of $F(x)$ can be expressed via the following $I/E$-rules, in analogy to 
the well-known rules for the transparent truth predicate.
\begin{equation}
\label{eq:ndfalsity}
\infer[(F I)]{F(\hat{\varphi})}{\neg \varphi}\ \ \ \ \ \ \ \ \
\infer[(F E)]{\neg \varphi}{F(\hat{\varphi})}
\end{equation}
Notably, those rules are {\em impure} in that they feature a connective (`$\neg$' here) different
from the one introduced/eliminated by the rules.

{\em But, what can be done in a more general setting, where no analog to classical negation
is present (or definable), to have a transparent falsity predicate?}
\subsection{Bivalent $l$-sequents and  transparent truth/falsity predicates}
\subsubsection{Bivalent l-sequents}
Consider now binary $l$-sequents $\Pi=\Gamma : \Delta$ 
(i.e., where $n=2$).
The advantage of this notation in the bivalent case is that it enables expressing falsity of a formula $\varphi$
{\em without appealing to negation}, just using a
located formula $(\varphi, f)$.
Note that both false assumptions and false conclusions are
allowed, residents of the respective  $\Gamma$ (assumptions)
and $\Delta$ (conclusions).

I consider a sound and complete  ND-system $\cN^{2}$ for the logic of bivalent  valid $l$-sequents.
Since the connectives are orthogonal to our current concerns,
I omit the presentation of their $I/E$-rules. However, this system
allows speaking proof-theoretically about my concerns.

The proof system $\cN^{2}$ is a special case of $\cN^{n}$ for $n=2$.
The general system is   presented in an  appendix.

We now can state that the falsity predicate $F(x)$ is disqoutational by  the following analogy to (\ref{eq:falsity}), without any appeal to negation.
\begin{equation}
\label{eq:psdft}
(PSD_{ft})\ 
\Gamma : \Delta, (F(\hat{\varphi}),t) \dashv\vdash_{\cN^{2}} 
\Gamma : \Delta, (\varphi,f)
\end{equation}
That is: if $F(\hat{\varphi })$ is true, indicated by its $t$-location 
of the l.h.s.,
then $\varphi$ is false, indicated by $f$-location
 of the r.h.s.,
and vice versa.
\begin{equation}
\label{eq:psdff}
(PSD_{tf})\ 
\Gamma : \Delta, (F(\hat{\varphi}),f) \dashv\vdash_{\cN^{2}} 
\Gamma : \Delta, (\varphi,t)  
\end{equation}
That is: if $F(\hat{\varphi })$ is false, indicated by its $f$-location 
of the l.h.s.,
then $\varphi$ is true, indicated by $t$-location
 of the r.h.s.,
and vice versa.

Note the use of a false conclusion in this formulation of the disquotation
property of the falsity predicate. This is how the use of (binary) $l$-sequents
overcomes the lack of direct means to refer to falsity without using (classical) negation.

Similarly, we can add to $\cN^{2}$ the following {\em pure} falsity transparency $I/E$-rules,
not appealing to `$\neg$':
\begin{equation}
\label{eq:IEF}
\infer[(F I_{t})]{\Gamma : \Delta, (F(\hat{\varphi}),t)}
{
\Gamma : \Delta, (\varphi,f)
}\ \ \ \ \ \ \
\infer[(F E_{t})]{\Gamma : \Delta, (\varphi,f)}
{
\Gamma : \Delta, (F(\hat{\varphi}),t)
}
\end{equation}
Again, for the ($FI_{t}$)-rule,  if $\varphi$ is false, indicated by its location $f$ in $\Delta$
of the premise,
then $F(\hat{\varphi})$ is true, indicated by $t$-locating it in 
$\Delta$ of the conclusion,
and similarly for the ($FE_{t}$)-rule. Note that in the formulation of these rules, 
both false assumptions and false conclusions are employed.
\begin{equation}
\infer[(F I_{f})]{\Gamma : \Delta, (F(\hat{\varphi}),f)}
{
\Gamma : \Delta,(\varphi,t) 
}\ \ \ \ \ \ \
\infer[(F E_{f})]{\Gamma : \Delta, (\varphi,t) }
{
\Gamma : \Delta, (F(\hat{\varphi}),f)
}
\end{equation}

Again, both (\ref{eq:psdft}) and (\ref{eq:psdff}) become derivable by means of the transparency 
$I/E$-rules for $F(x)$.

Next, those ideas are generalized for an arbitrary $n \ge 2$.
\section{Truth, falsity and their uniqueness}
\label{sec:unique}
In this section, I identify truth and falsity in $\cV$ and prove their uniqueness.
Recall that no other connectives besides $\cN_{i}$ are assumed to be present.
\subsection{Identifying truth}
\begin{definition}[truth]
A truth-value $v_{j} \in \cV$, for some $1 \le j \le n$, is a {\em truth} iff the following
holds for every $1\le i \le n$ and every $\varphi$:
\begin{equation}
(N_{t})\ 
\Gamma : \Delta, (N_{i}\varphi, j)\
\dashv\vdash_{\cN^{n}}\ 
\Gamma : \Delta,\{ (\varphi, k)\ |\ k \ne i \}
\end{equation}
\end{definition}
That is, for any $1 \le i \le n$, the locating $N_{i}\varphi$ at $v_{j}$ 
(i.e., at  a truth) is 
necessary and sufficient for locating $\varphi$ itself with 
$\{v_{k}\ |\ k \ne i\}$ (i.e., {\em not} with $v_{i}$).
Thus, being located with a truth assures the intended meaning of $N_{i}\varphi$
as {\em not assigning} $v_{i}$ to $\varphi$ (for all $i$s).

For this definition to make sense, I need to show that truth is unique; that is, if $v_{j}$ and
$v_{k}$ are truths, then $j=k$. The existence of a truth
is shown at the end of 
the paper, in \ref{eq:TT}.
\begin{proposition}[uniqueness of truth]
If both $v_{j}$ and $v_{k}$ are truths, then $j=k$.
\end{proposition}
{\bf Proof}: Assume, towards a contradiction, that for $j \ne k$ both $v_{j}$ and $v_{k}$ are truths. Then, 
\begin{equation}
\begin{array}{ccc}
(N_{k}\varphi,j) :  (N_{k}\varphi,j) &
(N_{t}, {\rm with}\  i=k)\over{\dashv\vdash_{\cN^{n}}} &( N_{k}\varphi,j)  :  \{(\varphi,m)\ |\ m \ne k\}\\ \\
 & (N_{t}, {\rm with}\  i=k)\over{\dashv\vdash_{\cN^{n}}} & (N_{k}\varphi,j) :  (N_{k}\varphi,k)
\end{array}
\end{equation}
But,
\begin{equation}
\label{eq:contradiction}
\infer[(cut)]{ : }
{\infer[(N_{t}\ {\rm with}\ i=j)]{:( N_{k}(N_{k}\varphi), j)}
{\infer[(shift)]{: \{(N_{k}\varphi, m)\ |\ m \ne j \}}
{\infer[(c_{j,k})]{(*)\ (N_{k}\varphi,j)  : }
{(N_{k}\varphi,j)  :  (N_{k}\varphi,j)
   &
 (N_{k}\varphi,j)  : (N_{k}\varphi,k)
}
}
}
  &
 \deduce{( N_{k}(N_{k}\varphi), j)}{{\rm substitute}\ N_{k}\varphi\ {\rm for}\ \varphi\ {\rm in}\ (*)} :
  }
\end{equation}
a contradiction.\newline
 Thus, $j=k$.\newline
For the coordination rule ($c_{j,k}$) and the ($cut$) rule -- see the appendix.

Since the numbering of the truth-values in $\cV$ is arbitrary, we assume henceforth that $v_{1}$ is the unique truth in $\cV$.
\subsection{Identifying falsity}
\begin{definition}[falsity]
A truth-value $v_{j} \in \cV$, for some $1 \le j \le n$, is a {\em falsity} iff the following
holds for every $1\le i \le n$ and every $\varphi$:
\begin{equation}
(N_{f})\ 
\Gamma : \Delta, (N_{i}\varphi,j)\
\dashv\vdash_{\cN^{n}}\ 
\Gamma :\Delta, (\varphi,i)
\end{equation}
\end{definition}
That is, for any $1 \le i \le n$, locating $N_{i}\varphi$ with  $v_{j}$  (i.e., a falsity) is 
necessary and sufficient for locating $\varphi$ itself with 
 $v_{i}$.
Thus, being located  at a falsity assures the intended meaning of $N_{i}\varphi$
as {\em not assigning} $v_{i}$ to $\varphi$ (for all $i$s) does not hold.

Again, for this definition to make sense, I need to show that falsity is unique; that is, if $v_{j}$ and
$v_{k}$ are falsities, then $j=k$.
The existence of a falsity is shown at the end of the paper, in 5.27.
\begin{proposition}[uniqueness of falsity]
If both $v_{j}$ and $v_{k}$ are falsities, then $j=k$.
\end{proposition}
{\bf Proof}: Assume, towards a contradiction, that for $j \ne k$ both $v_{j}$ and $v_{k}$ are falsities. Then, 
\begin{equation}
\begin{array}{ccc}
(N_{k}\varphi, j) : (N_{k}\varphi, j) &
(N_{f}, {\rm with}\  i=k)\over{\dashv\vdash_{\cN^{n}}} & (N_{k}\varphi, j) :  (\varphi,k)\\ \\
 & (N_{f}, {\rm with}\  i=k)\over{\dashv\vdash_{\cN^{n}}} & (N_{k}\varphi, j) :  (N_{k}\varphi,k)
\end{array}
\end{equation}
But,
\begin{equation}
\infer[(c_{j,k})]{(N_{k}\varphi, j) :}
{(N_{k}\varphi, j):  (N_{k}\varphi,j)
   &
(N_{k}\varphi, j) :  (N_{k}\varphi,k)
}
\end{equation}
A contradiction is now derived as in (\ref{eq:contradiction}).\newline
 Thus, $j=k$.

Since the numbering of the truth-values in $\cV$ is arbitrary, we assume henceforth that $v_{n}$ is the unique falsity in $\cV$.

\section{A natural deduction system for $N_{i}$}
I again assume that $N_{i},\ 1 \le i \le n$ are all the operators
in the object-language, ignoring at this point any other connectives.
\subsection{The rules for $N_{1}$}
Let us start with the case of $N_{1}$, with $N_{1} \varphi$ being true (i.e.,
having truth-value $v_{1} =  t$).
In this case, by the intended interpretation, $\varphi$ indeed does not have the truth-value
 $v_{1} =t$.\newline
The natural $I/E$-rules rules fitting the intended interpretation
are the following (cf. (\ref{eq:IEF})).
\begin{description}
\item[$I$-rule:] 
\begin{equation}
\infer[(N_{1} I_{1})]{\Gamma: \Delta, (N_{1}\varphi,1)}
{
\Gamma:  \Delta, \{(\varphi,j)\ | j \ne 1\}
}
\end{equation}
The premise expresses  that $\varphi$ has any of the
truth-values $v_{j}$, for $j \ne 1$, that is
$\varphi$ {\em having}  truth-value $v_{1}$,
{\em is not true}.
The conclusion is that $N_{1} \varphi$ is located at  $v_{1}$ (i.e., is true).
\item[$E$-rule:]
\begin{equation}
\infer[(N_{1} E_{1})]{\Gamma:  \Delta, \{(\varphi,j)\ | j \ne 1\}}
{
\Gamma: \Delta, (N_{1}\varphi,1)
}
\end{equation}
The premise expresses that $N_{1} \varphi$ is true, located in
 $v_{1}$.
The elimination is by distributing $\varphi$ itself, disjunctively, to all $v_{j},\ j \ne 1$.
\end{description}
Next, consider the situation where $N_{1} \varphi$ is false, i.e.,
having the truth-value $v_{n}$.
In this case, by the intended interpretation, {\em it is not the case}
that $\varphi$  does not have the truth-value
 $v_{1} =t$. In other words, $\varphi$ {\em has} the value
 $v_{1}$.\newline
The natural $I/E$-rules rules fitting the intended interpretation
are the following.
\begin{description}
\item[$I$-rule:]
\begin{equation}
\infer[(N_{1} I_{n})]{\Gamma : \Delta, (N_{1}\varphi, n)}
{
\Gamma : \Delta, (\varphi, 1)
}
\end{equation}
\item[$E$-rule:]
\begin{equation}
\infer[(N_{1} E_{n})]
{\Gamma : \Delta, (\varphi, 1)}
{\Gamma : \Delta, (N_{1}\varphi, n)}
\end{equation}
\end{description}

Next, suppose $1 <i \le n$, and suppose $N_{1} \varphi$ has truth value $v_{i}$.

{\bf A failing attempt}:\newline
To direct the thought, consider first $i=2$ and suppose that $v_{2}$, in some sense,
 means ``almost true''. What does it mean that it is ``almost true'' that $\varphi$
 does not have the truth-value $v_{1}=t$?\newline
A suggestive interpretation of this situation is that
 either $\varphi$ has {\em just one}  other truth-value $v_{j}$
 for $j \ne 1$, {\em or} it {\em does} have truth-value $v_{1}$. 
 
 Generalizing, it is suggestive to  interpret $\varphi$ not having truth-value $v_{1}=t$ to a truth degree $v_{i}$
 as either $\varphi$ having any other  truth-value $v_{j} \in A \subset \cV$, where $A$
 is of size $i-1$, {\em or} $\varphi$ {\em does} have the value $v_{1}$.\newline
 This would lead to the following $I/E$-rules:
\begin{description}
\item[$I$-rule:] 
For some $A \subset \hat{n}$ of size $i-1$, where $1 \not\in A$, there is a rule
\begin{equation}
\infer[(N_{1} I_{A}-attempted)]{\Gamma: \Delta, (N_{1}\varphi, i)}
{
\Gamma: \Delta, (\varphi,A \join \{1\})
}
\end{equation}
The premise expresses that $\varphi$ has one of the $i-1$ truth-values in $A$
(that exclude $v_{1}$), or does have truth-value $v_{1}$.
The conclusion locates $N_{1}\varphi$ in $v_{i}$.
\item[$E$-rule:] 
For every $A \subset \hat{n}$ of size $i-1$, where $1 \not\in A$, there is a rule
\begin{equation}
\infer[(N_{1} E_{A}-attempted)]{\Gamma: \Delta, (\varphi,A\join \{1\})}
{
\Gamma:\Delta, (N_{1}\varphi,i)
}
\end{equation}
The premise asserts that $N_{1}\varphi$ has truth-value $v_{i}$.
The conclusion distributes $\varphi$ disjunctively among the $i-1$  truth-values (excluding $v_{1}$), or in $v_{1}$.
\end{description}
Unfortunately, {\em this attempt fails}!\newline

Consider the following derivation.
\[
\infer[(N_{1} I_{A}-attempted)]{\Gamma : \Delta, (N_{1}\varphi, j),\ j \ne n}
{
  \infer[(W)]{\Gamma : \Delta, (\varphi, A \join \{1\})}
    {
     \infer[(N_{1}  E_{n})]{\Gamma : \Delta,(\varphi, 1)}
      {
       \Gamma : \Delta,(N_{1}\varphi, n)
      }
    }
}
\]
But by applying coordination to the assumption and conclusion
of the above derivation, we get
\[
\infer[(c_{j,n})]{\Gamma : \Delta}
{\Gamma : \Delta,(N_{1}\varphi, j,\ j \ne n)
 &
\Gamma : \Delta,(N_{1}\varphi, n)
}
\]
That is, $N_{1} \varphi$ ``disappeared''! This is, of course, wrong.

To understand what is going on and reach the correct rules,
consider again the informal interpretation of $N_{1}\varphi$:
it means negating {\em in the meta-language} that the truth-value of
$\varphi$ is $v_{1}$. However,
the meta-language employs classical logic, which is bivalent.
Recall that $N_{1} \varphi$ having truth-value $v_{i}$  means that, for a ``truth-degree'' $i$,
$\varphi$ does not have the truth-value $v_{1}$. So, the above 
interpretation must be either true or false. Thus, in the logic,
$N_{1}\varphi$ can only be located at $v_{1}$  (truth) or $v_{n}$
falsity. It cannot be located at any other $v_{j}$,\ $j\ne 1,n$.

This is reflected in $(N_{1}\varphi,j)$ having {\em no} I-rule,
and the following E-rule:
\begin{equation}
\infer[(N_{1} E_{j})]{\Gamma : \Delta}
{
\Gamma : \Delta, (N_{1}\varphi, j),\ j \ne 1,n
}
\end{equation}

\subsection{The general case $N_{k}$}
I now apply the same considerations to the general case $N_{k}$ for $1 < k \le n$.

\begin{description}
\item[$I$-rule:] 
\begin{equation}
\infer[(N_{k} I_{1})]{\Gamma: \Delta, (N_{k}\varphi,v_{1})}
{
\Gamma:  \Delta, \{(\varphi,j)\ | j \ne k\}
}
\end{equation}
The premise expresses  that $\varphi$ has any of the
truth-values $v_{j}$, for $j \ne k$, that is
$\varphi$ {\em having}  truth-value $v_{k}$,
{\em is not true}.
The conclusion is that $N_{k} \varphi$ is located at  $v_{1}$ (i.e., is true).
\item[$E$-rule:]
\begin{equation}
\infer[(N_{k} E_{1})]{\Gamma:  \Delta, \{(\varphi,j)\ | j \ne k\}}
{
\Gamma: \Delta, (N_{k}\varphi,v_{1})
}
\end{equation}
The premise expresses that $N_{1} \varphi$ is true, located in
 $v_{1}$.
The elimination is by distributing $\varphi$ itself, disjunctively, to all $v_{j},\ j \ne k$.
\end{description}
Next, consider the situation where $N_{k} \varphi$ is false, i.e.,
having the truth-value $v_{n}$.
In this case, by the intended interpretation, {\em it is not the case}
that $\varphi$  does not have the truth-value
 $v_{k} $. In other words, $\varphi$ {\em has} the value
 $v_{k}$.\newline
The natural $I/E$-rules rules fitting the intended interpretation
are the following.
\begin{description}
\item[$I$-rule:]
\begin{equation}
\infer[(N_{k} I_{n})]{\Gamma : \Delta, (N_{k}\varphi, n)}
{
\Gamma : \Delta, (\varphi, k)
}
\end{equation}
\item[$E$-rule:]
\begin{equation}
\infer[(N_{k} E_{n})]
{\Gamma : \Delta, (\varphi, k)}
{\Gamma : \Delta, (N_{k}\varphi, n)}
\end{equation}
\end{description}
Again, $N_{k} \varphi$ cannot have any other truth-value
 except $v_{1}$ or $v_{n}$. 
 This is again reflected in $(N_{k}\varphi)$ having {\em no} I-rule,
and the following E-rule:
\begin{equation}
\infer[(N_{k} E_{j})]{\Gamma : \Delta}
{
\Gamma : \Delta, (N_{k}\varphi, j),\ j \ne 1,n
}
\end{equation}
A somewhat tedious calculation can show that those $I/E$-rules are
generated, by the reciepe for operational rules in the appendix,
from the following truth-tables for the $N_{i}$s:
\begin{equation}
\label{eq:TT}
\begin{array}{c}
N_{i}(v_{j})=v_{1},\ {\rm for}\ j\ne i\\
N_{i}(v_{i})=v_{n}
\end{array}
\end{equation}
This establishes the existence of truth and falsity in the
general case.
\newpage
 {\bf  \large Appendix:  The proof-system $\cN^{n}$}
 \begin{description}
 \item[initial poly-sequents:]
 For every $1 \le i \le n$:
$\Gamma,(\varphi,i) : \Delta,(\varphi,i)$
\item[shifting rules:]
\[
\infer[(\overrightarrow{s}_{i})]{\Gamma : \Delta,\varphi\times \overline{i}}
{
\Gamma,(\varphi,i) : \Delta
}\ \ \ \ \ \
\infer[(\overleftarrow{s}_{i,j})]{\Gamma,(\varphi,j)
: \Delta}
{
 \Gamma :  \Delta,(\varphi,i)
}, j \ne i
\]
\item[coordination:]
\[
\infer[(c_{i,j})]{\Gamma :  \Delta}
{\Gamma :  \Delta,(\varphi,i)
 &
\Gamma :  \Delta,(\varphi,j)
},\ \ i \ne j
\]
From ($c_{i,j}$) the {\em Weakening} rules are derivable:
\[
\infer[(WL)]{\Gamma,\Gamma^{\prime}: \Delta}
{
\Gamma: \Delta
}
\ \ \ \ \ \
\infer[(WR)]{\Gamma: \Delta,\Delta^{\prime}}
{
\Gamma: \Delta
}
\]
\item[operational rules:] Those are irrelevant here, and are presented for completeness only.
The guiding lines for the construction are the following, expressed in terms of a
generic $p$-ary operator, say `$*$'.
\begin{description}
\item[($* I$):] Such rules introduce a conclusion 
$\Gamma :  \Delta,(*(\varphi_{1},\cdots,\varphi_{p}),k)$.
\begin{itemize}
\item
In general, if in the truth-table for `$*$' the values $v_{i_{j}}$ for $\varphi_{j}$, $1\le j \le p$, 
 yield the value $v_{k}$ for $*(\varphi_{1},\cdots,\varphi_{p})$, then there is a rule
\[
\label{eq:*I}
\infer[(* I_{i_{1},\cdots,i_{p},k})]{\Gamma :  \Delta,(*(\varphi_{1},\cdots,\varphi_{p}),k)}
{\{ \Gamma :  \Delta,(\varphi_{j},i_{j})\ |\  1 \le j \le p \}}
\]
The rule $(* I_{i_{1},\cdots,i_{p},k})$ has, thus, $p$ premises.
\end{itemize}
\item[($* E$):] Such rules have a major premise
$\Gamma : \Delta, (*(\varphi_{1},\cdots,\varphi_{p}), k)$.
\[
\infer[(* E_{k})]
{
\Gamma
:\Delta}
{
  \Gamma : \Delta, (*(\varphi_{1},\cdots,\varphi_{p}), k)
     &
\{ \Gamma ,*(\varphi_1 , k_1),\cdots,(\varphi_p , k_p)
: \Delta |
*(v_{k_1},\cdots,v_{k_p}) = v_k \}
}
\]
for each $k = 1,\cdots,n$.
\newline
\end{description}
 \end{description}
 A detailed discussion of this system, presented in a different but equivalent notation, 
  can be found in \cite{FK17a}.
  
  {\bf Acknowledgement}
  I thank Michael Kaminski for hos involvement in this paper.
\newpage
 \bibliography{pr_theory,mypapers,logic,philosophy}
\bibliographystyle{plain}
\end{document}